\documentstyle[12pt,axodraw,epsfig]{article}

\def\lsim{\:\raisebox{-0.5ex}{$\stackrel{\textstyle<}{\sim}$}\:}
\def\gsim{\:\raisebox{-0.5ex}{$\stackrel{\textstyle>}{\sim}$}\:}

\begin{document}
{\flushright
\hspace{30mm} hep-ph/0301203\\
\hspace{30mm} KIAS-P02032 \\
\hspace{30mm} UCCHEP/22-03 \\
\hspace{30mm} January 2003 \\
}
\vspace{1cm}

\begin{center}
{\Large\sc {\bf Searching for a light Fermiophobic Higgs Boson 
at the Tevatron}}
\vspace*{3mm}
\vspace{1cm}

{\large {Andrew G. Akeroyd}$^a$, {Marco A. D\'\i az}$^b$}
\vspace{1cm}

{\sl
a: Korea Institute for Advanced Study, 207-43 Cheongryangri-dong,\\
Dongdaemun-gu, Seoul 130-772, Republic of Korea \\
\vspace*{0.2cm}
b: Departamento de F\'\i sica, Universidad Cat\'olica de Chile,\\
Avenida Vicu\~na Mackenna 4860, Santiago, Chile 
}
\end{center}

\vspace{2cm}

\begin{abstract}
We propose new production mechanisms for light fermiophobic 
Higgs bosons ($h_f$) with suppressed couplings
to vector bosons ($V$) at the Fermilab Tevatron. These mechanisms 
(e.g. $qq'\to H^\pm h_f$) are complementary to 
the conventional process $qq'\to Vh_f$, which suffers from a strong
suppression of $1/\tan^2\beta$ in realistic models with a $h_f$.
The new mechanisms extend the coverage at the 
Tevatron Run II to the larger $\tan\beta$ region, and offer 
the possibility of observing new event topologies with up to 4 photons. 

\end{abstract}

\newpage
\section{Introduction}

The study of extensions of the Standard Model (SM) 
which include more than one Higgs doublet \cite{Gunion:1989we}
has received much attention in last 20 years.
The SM predicts one neutral Higgs scalar ($\phi^0$) with 
branching ratios (BRs) which are functions of $m_{\phi^0}$.
It is predicted to decay dominantly via $\phi^0\to b\overline b$
for $m_{\phi^0}\le 130$ GeV, and  $\phi^0\to VV^{(*)}$
(where $V=W^\pm,Z$) for $m_{\phi^0}\ge 130$ GeV.
The minimal extension of the SM contains an additional 
$SU(2)\times U(1)$ Higgs doublet, the ``Two Higgs Doublet
Model'' (2HDM), and the resulting particle spectrum consists of
2 charged Higgs bosons $H^+$, $H^-$ and 3 neutral members $h^0$, 
$H^0$ and $A^0$.
Assuming that each fermion type (up,down) couples to only one Higgs doublet
\cite{Glashow:1976nt}, which
eliminates tree-level Higgs mediated flavour changing neutral currents,
leads to 4 distinct versions of the 2HDM
\cite{Barger:1989fj}. Due to the increased parameter 
content of the 2HDM 
the BRs of the neutral Higgs bosons may be significantly different
to those of $\phi^0$ \cite{Gunion:1989we},\cite{Akeroyd:1996di}.
In recent years LEP2 has carried out searches \cite{Abbiendi:2000ug}
for such Higgs bosons with enhanced BRs to lighter 
fermions and bosons (e.g. $c\overline c, \tau^+\tau^-,gg$). 
The phenomena known as ``fermiophobia'' \cite{Weiler:1987an}
which signifies very suppressed or zero coupling to the fermions,
may arise in a particular version of the 2HDM called type I 
\cite{Haber:1979jt}.
Such a fermiophobic Higgs ($h_f$)\cite{Barger:1992ty,Pois:1993ay,
Stange:1994ya,Diaz:1994pk,Akeroyd:1996hg,Barroso:1999bf,Brucher:1999tx}  
would decay dominantly
to two bosons, either $h_f\to \gamma\gamma$ (for $m_{h_f}\le 90$ GeV)
or $h_f\to VV^{(*)}$ for ($m_{h_f}\ge 90$ GeV) 
\cite{Stange:1994ya,Diaz:1994pk}. 
This would give a very clear experimental 
signature, and observation of such a particle would strongly 
constrain the possible choices of the underlying Higgs sector.

Fermiophobic Higgs bosons have been searched for actively at LEP 
and the Tevatron.  All four collaborations at LEP 
(OPAL\cite{Abbiendi:2002yc}, DELPHI\cite{Abreu:2001ib},
ALEPH\cite{Heister:2002ub}, L3\cite{Achard:2002jh}) utilized the 
channel $e^+e^-\to h_fZ$, $h_f\to \gamma\gamma$
and obtained lower bounds of the order $m_{h_f}\ge 100$ GeV.
L3 \cite{Mans:ji} is the only collaboration yet to consider
$h_f\to WW^*$ decays. OPAL \cite{Abbiendi:2002yc} and DELPHI
\cite{Abreu:2001ib} also searched in the channel $e^+e^-\to H_FA^0$, 
$H_F\to \gamma\gamma$.
In run I at the Tevatron the mechanism $qq'\to V^*\to h_fV$,$h_f\to 
\gamma\gamma$
was used, with the dominant contribution coming from $V=W^\pm$.
The limits on $m_{h_f}$ from the D0 and CDF collaborations 
are respectively 78.5 GeV \cite{Abbott:1998vv}
and 82 GeV \cite{Affolder:2001hx} at $95\%$ $c.l$.
Run II will extend the coverage of $m_{h_f}$ beyond that of LEP.

However, all these mass limits assume that the $h_fVV$ coupling is of 
the same strength as the SM coupling 
$\phi^0VV$, which in general would not be the 
case for a $h_f$ in a realistic model e.g. the 2HDM (type~I) or
the Higgs triplet model of \cite{Georgi:1985nv}, \cite{Akeroyd:1998zr}.  
Therefore one could imagine the scenario of a very light $h_f$ 
($m_{h_f}<< 100$ GeV) which has eluded the current searches at LEP and the 
Tevatron Run I due to suppression in the coupling $h_fVV$.
Such a $h_f$ could also escape detection in the Tevatron Run II. 
In this paper we propose new production 
mechanisms at the Tevatron Run II 
which are effective even when the coupling $h_fVV$ is very suppressed.

Our work is organized as follows. Section 2 gives an introduction
to the phenomenology of fermiophobic Higgs bosons while
Section 3 presents the new production mechanisms.
Section 4 contains our numerical results with conclusions in section 5.

\section{Models with Fermiophobia}

A fermiophobic Higgs boson ($h_f$) may arise in a 2HDM in which
one $SU(2)\times U(1)$ Higgs doublet ($\Phi_2$) 
couples to all fermion types,
while the other doublet ($\Phi_1$) 
does not. Both doublets couple to the gauge 
bosons via the kinetic term in the Lagrangian. One vacuum
expectation value ($v_2$) gives mass to all fermion types, while
gauge bosons receive mass from both $v_1$ and $v_2$. This model
(usually called ``Type I'') was first proposed in 
\cite{Haber:1979jt}. Due to the mixing in the
CP--even neutral Higgs mass matrix (which is diagonalized by $\alpha$)
both CP--even eigenstates $h^0$ and $H^0$ can couple to the fermions.
The fermionic couplings of the lightest CP--even Higgs 
$h^0$ take the form 
\begin{equation}
h^0f\overline f \sim \cos\alpha/\sin\beta
\end{equation}
where $f$ is any fermion, and $\beta$ is defined by $\tan\beta=v_2/v_1$.

Small values of $\cos\alpha$ would seriously suppress the fermionic 
couplings, 
and in the limit $\cos\alpha \to 0$ the coupling $h^0f\overline f$ would
vanish at tree--level, giving rise to   
fermiophobia (sometimes called a ``bosonic'' or ``bosophillic'' Higgs):
\begin{center}
\vspace{-30pt} \hfill \\
\begin{picture}(120,70)(0,23) % y_2 controls equation position
\DashLine(30,30)(60,30){3}
\Text(20,30)[]{$h_f$}
\ArrowLine(60,30)(90,60)
\Text(100,60)[]{$f$}
\ArrowLine(90,0)(60,30)
\Text(100,0)[]{$\overline f$}
\end{picture}  %after this line goes the equation
$\sim \, 0$
\vspace{30pt} \hfill \\
\end{center}
\vspace{10pt}
However, at the 1--loop level there will be an effective vertex
$h_ff\overline f$ mediated by loops involving vector bosons and other
Higgs particles. These loop contributions are infinite and a counterterm is 
necessary to renormalize it. The counterterm is fixed with an experimental 
input, leading to an arbitrariness in the definition of the tree level 
vertex, or equivalently, in the mixing angle $\alpha$ \cite{Diaz:1994pk}.
It is customary to define an extreme fermiophobia, where $h_f$ remains 
fermiophobic to the 1-loop level/all orders with branching ratios given by 
\cite{Stange:1994ya},\cite{Diaz:1994pk}. In general, one would
expect some (small) coupling to fermions, from both 
tree--level diagrams and one loop diagrams.
\begin{center}
\vspace{-30pt} \hfill \\
\begin{picture}(120,70)(0,23) % y_2 controls equation position
\DashLine(30,30)(60,30){3}
\Text(20,30)[]{$h_f$}
\ArrowLine(60,30)(90,60)
\Text(100,60)[]{$f$}
\ArrowLine(90,0)(60,30)
\Text(100,0)[]{$\overline f$}
\GCirc(60,30){10}{0.5}
\end{picture}  %after this line goes the equation
$\sim \, 0$
\vspace{30pt} \hfill \\
\end{center}
\vspace{10pt}

The Higgs Triplet
model (HTM) discussed in \cite{Georgi:1985nv},\cite{Akeroyd:1998zr}
is another possible origin for a $h_f$. In such models gauge invariance
forbids the tree--level coupling of some triplet Higgs bosons to fermions, 
and so suppressed BRs to fermions are expected without requiring specific 
mixing angles. 

The main decay modes of a fermiophobic Higgs are:
\begin{center}
\vspace{-30pt} \hfill \\
\begin{picture}(200,70)(0,23) % y_2 controls equation position
%
% Left graph
%
\DashLine(30,30)(60,30){3}
\Text(20,30)[]{$h_f$}
\Photon(60,30)(90,60){3}{6.5}
\Text(120,60)[]{$\gamma,W^*,Z^*$}
\Photon(90,0)(60,30){3}{6.5}
\Text(120,0)[]{$\gamma,W^*,Z^*$}
\end{picture}
\vspace{30pt} \hfill \\
\end{center}
\vspace{10pt}
$h_f\to \gamma\gamma$ is the dominant decay for $m_{h_f}\lsim 95$ GeV
(sometimes called a ``photonic Higgs''),
with a BR near 100\% for $m_{h_f}\lsim 80$ GeV, decreasing to 
50\% at $m_{h_f}\approx 95$ GeV and to 1\% at $m_{h_f}\approx 145$ GeV.
In contrast, BR$(\phi^0\to \gamma\gamma)\approx 0.22\%$ is the
largest value in the SM for the two photon decay. In this paper
we shall be focusing on the possibility of a light 
($m_{h_f}\le 100$ GeV) for which the photonic decay mode always 
has a large BR.

BR$(h_f\to WW^*)$ supercedes BR$(h_f\to \gamma\gamma)$ when 
$m_{h_f}\gsim 95$ 
GeV, with a BR approaching $100\%$ for $110$ GeV $< m_{h_f}< 170$ GeV, and 
stabilizing at $\sim 70\%$ for $m_{h_f}\ge 2M_Z$.  
The decay $h_f\to ZZ^*$ is always subdominant, but for 
$m_{h_f}\ge 2M_Z$ 
approaches $30\%$. 
Recently, L3 \cite{Mans:ji} has included these $VV^*$ decays 
in their searches, and the discovery prospects of this decay mode
at the Tevatron Run II have been presented in \cite{Han:1998sp}.

%This BR decreases with 
%decreasing $M_F$ to 50\% at $M_F\approx 95$ GeV and to 1\% at 
%$M_F\approx 45$ GeV. 
%The decay $h\f\to Z^*Z^*$ also increases with $M_F$ 
%from 1\% at $M_F\approx 70$ GeV up to near 10\% for $M_F\approx 140$ GeV.
%The presence of the $WW$ threshold pushes $B(H_F\to Z^*Z^*)$ down to near 
%1\% at $M_F\approx 170$ GeV, but after that, the $ZZ$ threshold opens up
%reaching $B(H_F\to Z^*Z^*)=10\%$ again at $M_F\approx 190$ GeV.

Apart from the 2HDM (Type I) and the HTM, there are other models 
beyond the SM which allow the possibility of
a neutral Higgs boson with an enhanced BR to $\gamma\gamma$, as explained in 
\cite{Mrenna:2001qh}. These include $h^0$ of the MSSM, and $h^0$ of
top--condensate models. We will not consider these
models, which have a smaller BR$(h^0\to \gamma\gamma)$
than the fermiophobic models, and instead
focus on the 2HDM (Type I)\footnote{Another interesting 
possibility for a light Higgs boson with
enhanced decays to $\gamma\gamma$ has been considered in 
\cite{Dobrescu:2000jt},\cite{Larios:2001ma}. 
Here if $A^0$ is extremely light ($\le 0.2$ GeV) then 
BR$(A^0\to \gamma\gamma$) may be large.}.
Our results can also be quite easily extrapolated to the case of the HTM.

The conventional production mechanism for a $h_f$ at
$e^+e^-$ colliders is $e^+e^-\to Z^*\to h_fZ$, and at Hadron colliders
$qq'\to V^*\to h_fV$. Note that the gluon-gluon fusion mechanism
(via heavy quark loops) is not relevant for a $h_f$.
In the 2HDM (Type~I), the condition for tree--level fermiophobia
($\cos\alpha\to 0$) causes the coupling $h_fVV$ to be suppressed by
a factor 
\begin{equation}
h_fVV\sim \sin^2(\beta-\alpha)\to \cos^2\beta\;\; \equiv 1/(1+\tan^2\beta)
\end{equation}
Taking $\tan\beta\ge 3(10)$ implies a strong suppression 
of $\le 0.1 (\le 0.01)$ for the coupling $h_fVV$
with respect to the coupling $\phi^0VV$. 
This suppression is always possible for the lightest 
CP--even neutral Higgs in any of the 
4 types of the 2HDM  \cite{Gunion:1989we} and also occurs 
for the $h_f$ in the HTM \cite{Akeroyd:1998zr}. 
Therefore one can imagine the scenario of a very light $h_f$ which 
has eluded the searches via the mechanisms $e^+e^-/qq'\to h_f V$.
The possibility of a light 
$h_f$ has been known for a long time \cite{Diaz:1994pk} and
has been emphasized in \cite{Akeroyd:1996hg},\cite{Barroso:1999bf}.
LEP ruled out regions of the plane
[$m_{h_f},R\times BR(h_f\to \gamma\gamma)$], where $R$ is defined by:
\begin{equation}
R={\sigma(e^+e^-\to Zh_f) \over \sigma(e^+e^-\to Z\phi^0)}
\end{equation}

In a benchmark scenario of $R=1$, and assuming 
BR($h_f\to \gamma\gamma$) given by \cite{Stange:1994ya},
\cite{Diaz:1994pk}, each collaboration
derived a limit of around $m_{h_f}\ge 100$ GeV
\cite{Abbiendi:2002yc},\cite{Abreu:2001ib},
\cite{Heister:2002ub},\cite{Achard:2002jh}, with the 
combined LEP working group limit being $m_{h_f}\ge 109$ GeV \cite{Mans:ji}.
From the LEP plots it is trivial to see the necessary suppression 
in $R$ which would permit a light $h_f$ of a given mass, e.g. 
$m_{h_f}\le 80$ GeV (50 GeV)
requires $R\le 0.1 (0.01)$, which corresponds to $\tan\beta\ge 3 (10)$
in the 2HDM (Type I). Therefore sizeable regions of the 
[$m_{h_f},R\times BR(h_f\to \gamma\gamma)$] plane remain unexcluded for
small $R$ and small $m_{h_f}$.
OPAL \cite{Abbiendi:2002yc} also performed a search which is
sensitive to the production mechanism $e^+e^-\to h_fA^0$. 
This process ($\sim \sin^2\beta$ in the fermiophobic limit) is
complementary to $e^+e^-\to h_fZ$ ($\sim \cos^2\beta$). Therefore
the condition $m_{h_f}+m_A\ge \sqrt s$ must also be satisfied
in order for a light $h_f$ to escape detection at LEP2.

With the closure of LEP, the Tevatron Run II will continue the
search for $h_f$. Run II will use the same mechanism as
Run I ($qq'\to V^*\to Vh_f$) but has the advantage of a much increased
luminosity. Ref.\cite{Mrenna:2001qh} has shown that (for $R=1$)
$m_{h_f}$ can be discovered
(at $5\sigma$) up to 114 GeV (128 GeV) with 2 fb$^{-1}$ 
(30 fb$^{-1}$), which is an
improvement over the LEP limits. Similar conclusions were reached in
\cite{Landsberg:2000ht}. However, with the expected suppression
in the $h_fVV$ coupling ($R< 1$), $m_{h_f}\le 80$ GeV could still escape
detection. The aim of this paper is to show that other production mechanisms 
are available at the Tevatron Run II, and allow discovery of a $h_f$
even in the region where the process $qq'\to Wh_f$ is suppressed.

We will be using the most general (CP conserving) 2HDM potential 
\cite{Gunion:1989we}. This
potential is parametrized by 7 independent variables, which may be
taken as the four Higgs masses, two mixing angles ($\alpha,\beta$), and
a real quartic coupling ($\lambda_5$). 
\begin{equation}
V(\Phi_{1}, \Phi_{2})=V_{sym}+V_{soft}
\end{equation}
where
\begin{eqnarray}
 V_{sym} =  -\mu^2_{1}\Phi^{\dagger}_{1}\Phi_1
-\mu^2_{2}\Phi^{\dagger}_{2}\Phi_2+
\lambda_1(\Phi^{\dagger}_{1}\Phi_1)^2+ 
\lambda_2(\Phi^{\dagger}_{2}\Phi_2)^2+ \nonumber \\
\lambda_3(\Phi^{\dagger}_{1}\Phi_1)(\Phi^{\dagger}_{2}\Phi_2)
+\lambda_4|\Phi^{\dagger}_1\Phi_2|^2+
{1\over 2}[\lambda_5(\Phi^{\dagger}_1\Phi_2)^2+h.c]
\end{eqnarray}
and
\begin{equation}
V_{soft}=-\mu^2_{12}\Phi_1^{\dagger}\Phi_2+h.c
\end{equation}

The condition for tree-level fermiophobia corresponds to 
$\cos\alpha\to 0$, with $\alpha$ being an independent parameter.
Ref.~\cite{Barroso:1999bf} considered the fermiophobic limit 
in the context of two 6 parameter 2HDM potentials 
($V_A$ and $V_B$). In Ref.~\cite{Barroso:1999bf} the angle 
$\alpha$ is not a free parameter,
and the condition $\cos\alpha\to 0$ requires certain relations among 
the Higgs masses to be fulfilled. We shall take all the Higgs masses
as free parameters and set $\cos\alpha=0$, which guarantees tree-level
fermiophobia.

\section{Production Processes}

In this section we introduce the production processes which may offer
sizeable rates for $h_f$ in the region where the coupling
$h_fVV$ is very suppressed. These production mechanisms make use of 
the cascade decays $H^{\pm}\to h_fW^{(*)}$ or $A^0\to h_fZ^{(*)}$
which may have large BRs in the 2HDM (Type I) \cite{Akeroyd:1998dt}
and the HTM \cite{Akeroyd:1998zr}. 
These large BRs arise since the coupling of $H^{\pm}$ and $A^0$ to all the
fermions scales as $1/\tan\beta$, and thus for moderate to large $\tan\beta$
even the 3--body decays (i.e. with $V^*$) can have sizeable or dominant BRs.
We note that in the MSSM such decays (with $h_f$ replaced by $h^0$)
never attain very large BRs since $H^{\pm}$ and $A^0$ couple to the
down type fermions with strength $\tan\beta$. In addition, the decays 
$H^{\pm}\to h^0W^{(*)}$ or $A^0\to h^0Z^{(*)}$
are proportional to $\cos^2(\beta-\alpha)$ which is suppressed in a large
part of the MSSM parameter space, but (in contrast) is maximized in 
the parameter of $h_f$ with suppressed coupling to vector bosons. 

Refs.\cite{Mrenna:2001qh},\cite{Landsberg:2000ht}
considered two signatures from the $qq'\to WH_F$ mechanism, i)
inclusive $\gamma\gamma$ and ii) exclusive $\gamma\gamma V$.
The latter gives a better signal to background ratio 
and we will see that the cascade decay produces the 
necessary vector boson for the $\gamma\gamma V$ signature.

Below we list four production processes which are complementary
to the standard $qq'\to WH_F$ mechanism. They all make use of the
Higgs-Higgs-Vector boson coupling ($g_{HHV}$) which is either 
proportional to $\sin\beta$ (in the fermiophobic limit)
or independent of mixing angles (see Table 1). 
All mechanisms can offer non-negligible cross--sections in the large
$\tan\beta$ region. Moreover, double $H_F$ production can occur, 
resulting in distinctive $\gamma\gamma\gamma\gamma$ topologies. 

\begin{table}[htb]
\centering
\begin{tabular} {|c|c|c|c|} \hline
 & $H^{\pm}AW^{\mp}$ & $H^{\pm}h_fW^{\pm}$ & $h_fAZ$  \\ \hline
 $g_{HHV}$ & 1 & $\sin\beta$ & $\sin\beta$  \\ \hline
\end{tabular}
\caption{Mixing angle dependence of the couplings $H_iH_jV$}
\end{table}

\begin{itemize}
\item[{(i)}] $q\overline q\to \gamma^*,Z^*\to H^+H^-$: Quark anti-quark 
pair annihilation produces a pair of charged Higgs bosons via an intermediate 
photon or $Z$ boson in the s-channel:
\begin{center}
\vspace{-40pt} \hfill \\
\hspace{1cm}
\begin{picture}(350,70)(0,25) % y_2 controls equation position
\Photon(60,25)(118,25){4}{8}
\ArrowLine(60,25)(10,55)
\ArrowLine(10,-5)(60,25)
\DashLine(168,-5)(118,25){3}
\DashLine(118,25)(168,55){3}
\Text(2,55)[]{$\overline q$}
\Text(2,-5)[]{$q$}
\Text(179,-2)[]{$H^-$}
\Text(179,55)[]{$H^+$}
\Text(90,38)[]{$\gamma,Z$}
\DashLine(250,30)(280,30){3}
\Text(240,30)[]{$H^+$}
\DashLine(280,30)(310,60){3}
\Text(325,60)[]{$h_f$}
\Photon(310,0)(280,30){3}{6.5}
\Text(325,0)[]{$W^*$}
\end{picture}
\end{center}
\vspace{1cm}
The subsequent decay $H^{\pm}\to h_fW^{*}$ may provide two $W^*$ and two 
$h_f$, resulting in a distinctive $\gamma\gamma\gamma\gamma$
plus four fermion signal.

\item[{(ii)}] $qq'\to W^*\to H^{\pm}h_f$: Quark anti-quark annihilation 
into an intermediate $W$ boson producing a $h_f$ in association with a 
charged Higgs:
\begin{center}
\vspace{-40pt} \hfill \\
\hspace{1cm}
\begin{picture}(200,70)(0,25) % y_2 controls equation position
\Photon(60,25)(118,25){4}{8}
\ArrowLine(60,25)(10,55)
\ArrowLine(10,-5)(60,25)
\DashLine(168,-5)(118,25){3}
\DashLine(118,25)(168,55){3}
\Text(2,55)[]{$\overline q'$}
\Text(2,-5)[]{$q$}
\Text(179,-2)[]{$h_f$}
\Text(179,55)[]{$H^{\pm}$}
\Text(90,38)[]{$W^{\pm}$}
\end{picture}
\end{center}
\vspace{1cm}
This mechanism was covered in the case of the MSSM in
\cite{Djouadi:1999rc}, but only for the heavier CP--even
$H^0$. The rate for the lighter CP--even $h^0$ is 
suppressed by $\cos^2(\beta-\alpha)$, which is
small in a large region of the MSSM parameter space.
The cross--sections for $H^+h_f$ and $H^-h_f$ 
are identical, and will be summed over in our numerical analysis.
This process is phase space favoured over (i) and provides direct 
production of $h_f$. A vector boson ($W^*$) is provided by 
the decay $H^{\pm}\to W^{*}h_f$. 
In this way, double $h_f$ production occurs with a signature of 
$\gamma\gamma\gamma\gamma$ plus $V^*$.

\item[{(iii)}]  $qq'\to W^*\to H^{\pm}A^0$: Quark anti-quark annihilation
into an intermediate $W$ producing a charged Higgs in association with a 
CP--odd neutral Higgs:
\begin{center}
\vspace{-40pt} \hfill \\
\hspace{1cm}
\begin{picture}(350,70)(0,25) % y_2 controls equation position
\Photon(60,25)(118,25){4}{8}
\ArrowLine(60,25)(10,55)
\ArrowLine(10,-5)(60,25)
\DashLine(168,-5)(118,25){3}
\DashLine(118,25)(168,55){3}
\Text(2,55)[]{$\overline q'$}
\Text(2,-5)[]{$q$}
\Text(179,-2)[]{$A$}
\Text(179,55)[]{$H^{\pm}$}
\Text(90,38)[]{$W$}
\DashLine(250,30)(280,30){3}
\Text(240,30)[]{$A$}
\DashLine(280,30)(310,60){3}
\Text(325,60)[]{$h_f$}
\Photon(310,0)(280,30){3}{6.5}
\Text(325,0)[]{$Z^*$}
\end{picture}
\end{center}
\vspace{1cm}
This process is similar to (i) since no fermiophobic Higgs is produced 
directly. We sum over the rates for $H^+A^0$ and $H^-A^0$ as in (ii).
The decay $H^{\pm}\to h_fW^{*}$ or $A^0\to Z^*h_f$ provides a
gauge boson $V$ and a $h_f$. Again, double $h_f$ production may occur
giving rise to a final state of $\gamma\gamma\gamma\gamma$
$V^*V^*$. This mechanism
was considered in the context of the MSSM in \cite{Kanemura:2002hz}.

\item[{(iv)}] $q\overline q\to Z^* \to A^0h_f$: Quark anti-quark pair 
annihilation into an intermediate $Z$ producing a fermiophobic Higgs in 
association with a CP--odd neutral Higgs:
\begin{center}
\vspace{-40pt} \hfill \\
\hspace{1cm}
\begin{picture}(200,70)(0,25) % y_2 controls equation position
\Photon(60,25)(118,25){4}{8}
\ArrowLine(60,25)(10,55)
\ArrowLine(10,-5)(60,25)
\DashLine(168,-5)(118,25){3}
\DashLine(118,25)(168,55){3}
\Text(2,55)[]{$\overline q$}
\Text(2,-5)[]{$q$}
\Text(179,-2)[]{$h_f$}
\Text(179,55)[]{$A$}
\Text(90,38)[]{$Z$}
\end{picture}
\end{center}
\vspace{1cm}
This process is similar to (ii) and gives direct production of $h_f$ 
with a $Z$ boson arising from the decay $A^0\to h_fZ^{*}$. The 
$\gamma\gamma\gamma\gamma$ signal is also possible with 
this mechanism.

\end{itemize} 

Mechanisms i) and iv) are the hadron collider analogies of the LEP production
processes $e^+e^-\to H^+H^-$ and $e^+e^-\to A^0h_f$, but have the advantage 
of the larger $\sqrt s$ at the Tevatron. Mechanisms ii) and iii) are 
exclusive to a hadron collider. The cross-section formulae for all the 
processes can be found in \cite{Eichten:1984eu},\cite{Dawson:1998py}.
One may write a generic formula for (ii),(iii) and (iv):
\begin{equation}
\sigma(q\overline q\to H_iH_j)={G_F^2M_Z^4\over 96\pi \hat s}
g_{HHV}^2(v_q^2+a_q^2){\lambda^{3/2}\over (1-M^2_V/\hat s)^2}
\label{crosssection}
\end{equation}
where $H_i$,$H_j$ (with mass $M_i,M_j$) refer to any of the Higgs bosons 
$H^{\pm}$, $A^0$, $h_f$, $\lambda(M_i,M_j)$ is the usual two body phase 
space function, and $\hat s$ is the centre of mass energy for the partonic 
collision. In eq.~(\ref{crosssection}), $v_q$ and $a_q$ represent the vector 
and axial vector couplings of the incoming quarks to the vector boson 
mediating the process, and are given in Table 2.
In the same formula, $g_{HHV}$ is the Higgs--Higgs--Vector Boson coupling 
%in units of the $\phi^0ZZ$ coupling, 
which are listed in Table 1.

\begin{table}[htb]
\centering
\begin{tabular} {|c|c|c|} \hline
 & $Z$ & $W^{\pm}$   \\ \hline
 $v_u$ & $0.25-{2\over 3}\sin^2\theta_w$& $\sqrt 2\cos^2\theta_w$ \\ \hline
 $a_u$ & 0.25 & $\sqrt 2/\cos^2\theta_w$  \\ \hline 
 $v_d$ & $-0.25-{1\over 3}\sin^2\theta_w$ & $\sqrt 2\cos^2\theta_w$\\ \hline
 $a_d$ & -0.25 & $\sqrt 2/\cos^2\theta_w$   \\ \hline 
\end{tabular}
\caption{Values for $v_q$ and $a_q$}
\end{table}

\section{Numerical Results}

We now outline the calculation of the cross--section for the processes
(i)$\to$ (iv) under consideration. The partonic cross--sections
are given by eq.~(\ref{crosssection}).
These must then be scaled up to a $p\overline p$ cross--section.
In the partonic centre of mass system, the kinematic is defined as:
\begin{eqnarray}
& & \hat s =  (p_{q}+p_{q'})^2 = (p_{H_i}+p_{H_j})^2  \nonumber\\
& & \hat t =
\frac{1}{2}(M_{H_i}^2  + M_{H_j}^2) -
\frac{\hat s}{2}
+\frac{\hat s}{2} \kappa \cos\theta  \nonumber\\
& & \hat u = \frac{1}{2} (M_{H_i}^2  + M_{H_j}^2)-
\frac{\hat s}{2}-
\frac{\hat s}{2} \kappa \cos\theta \nonumber \\
& & \hat s+\hat t+\hat u = M_{H_i}^2 + M_{H_j}^2  \nonumber
\end{eqnarray}
Here $\kappa ^2=(\hat s-(M_{H_i} + M_{H_j})^2)(\hat s-(M_{H_i} -
M_{H_j})^2)/\hat s^2$. The hadronic cross--section for the process 
$p\overline p\to qq'\to H_iH_j$ can be expressed as follows:
\begin{equation}
\sigma(p\overline p\to qq'\to H_iH_j) = \int_{(M_{H_i}+M_{H_j})^2/s}^1 d\tau
\frac{d{\cal L}^{q\overline q}}{d\tau} \hat\sigma (\hat s = \tau s) \; .
\end{equation}
In the case of the Tevatron Run II $\sqrt s=2$ TeV.
\begin{equation} 
\frac{d{\cal L}^{qq'}}{d\tau}=\int_{\tau}^1\frac{dx}{x}f_q
(x;Q^2)f_{q'}(\tau/x;Q^2)
\end{equation}
where $\tau=x_1x_2$, with $x_1$ and $x_2$ being the momentum fraction
carried by each incoming parton. The parton distributions 
$f_q$ and $f_{q'}$ shall be taken at the typical scale $Q\approx M_{H_i}$. 
We shall be using the MRST2002 set from \cite{Martin:2001es}.
Note that QCD corrections increase the tree--level cross--section by
a factor of around 1.3 \cite{Dawson:1998py}. In our analysis we shall 
present results using the tree--level formulae only. For the
BRs of the fermiophobic Higgs we also work at tree--level 
and set $\cos\alpha=0$ to ensure exact fermiophobia.

The four new production mechanisms under consideration 
are generally expected to be ineffective for 
searches where the Higgs boson decays to quarks, since
backgrounds will be sizeable. However, in the case of $h_f$
we will show that they offer promising detection prospects
despite the moderate cross--sections. This is because the efficiency for the 
$\gamma\gamma V$ channel is high ($\approx 25\%$) \cite{Mrenna:2001qh}, 
and the decays $H^{\pm}\to h_fW^{*}$ or $A^0\to Z^*h_f$ may have 
very large BRs in the 2HDM (Model~I) discussed here. 
In much of the parameter space of interest 
($\tan\beta\ge 1$ and $m_{h_f} < 100$ GeV),
BR($H^{\pm}\to h_fW^{(*)}$) and BR($A^0\to Z^{(*)}h_f)$ are 
close to $100\%$. 
Hence a light $h_f$ can be produced in a cascade with almost 
negligible BR suppression (see \cite{Akeroyd:1998dt} for a 
quantitative analysis of these BRs). 
The cascade decays provide distinctive $\gamma\gamma\gamma\gamma$ 
signatures from all four mechanisms.
In our numerical analysis we will vary $m_{h_f}$ with particular
emphasis on $m_{h_f} < 100$ GeV. We will take 
$m_{H^\pm}\ge 90$ GeV (roughly the lower bound from LEP2)
and $M_A$ is constrained by
$m_A+m_{h_f}\ge 200$ GeV from negative searches in the channel
$e^+e^-\to h_fA^0$. 
%Thus we expect mechanisms involving $H^\pm$
%to have the largest cross--sections. 
For the expected 2 fb$^{-1}$
of data from Run IIa, which might be available by 2005/2006,
we assume a threshold of observability of 10 fb for the cross--sections.
Larger data samples of up to 15 fb$^{-1}$ would require even smaller values.

\begin{figure}
\centerline{\protect\hbox{\epsfig{file=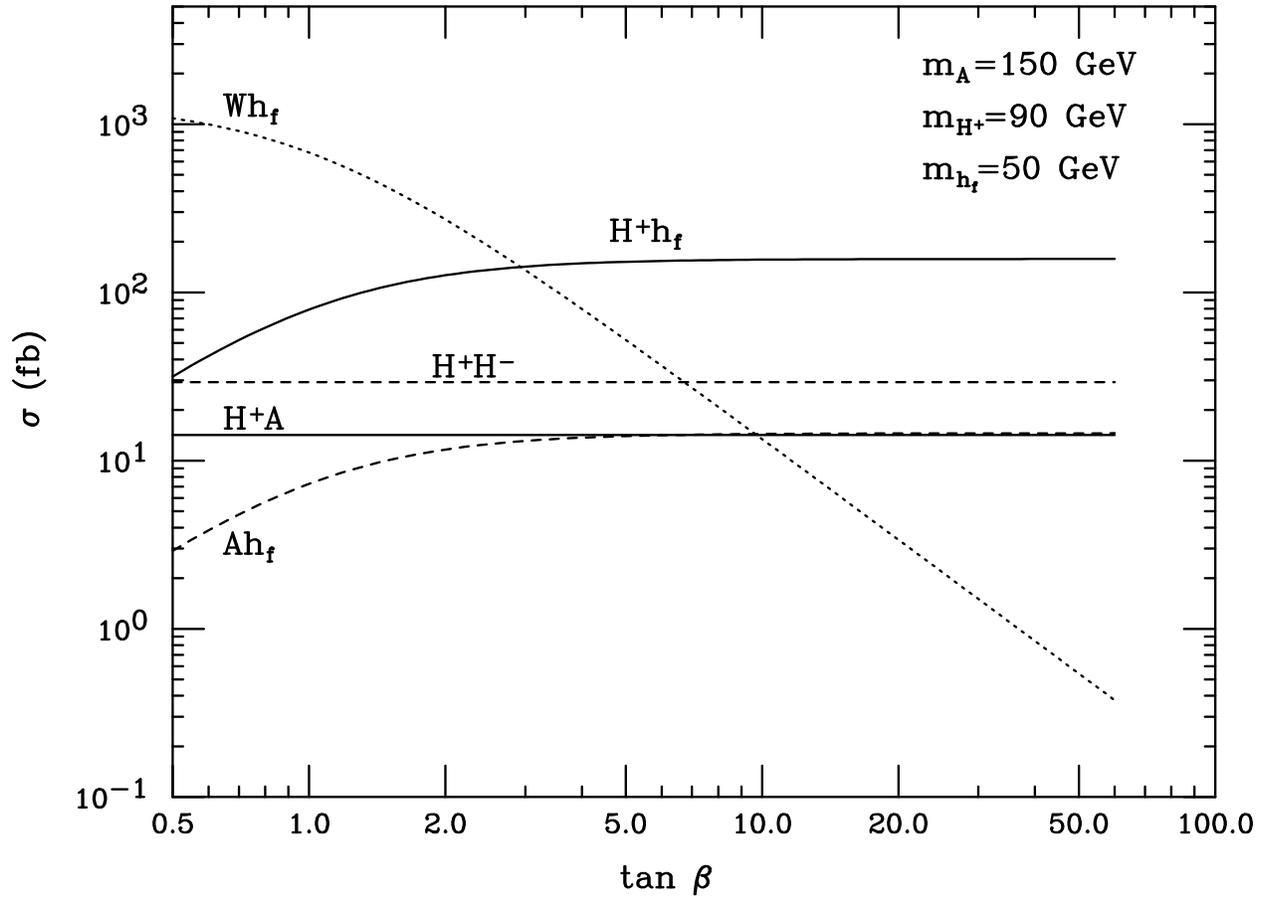,width=0.75\textwidth,angle=90}}}
\caption{\it Production cross--section of five different modes leading to 
a fermiophobic Higgs boson as a function of $\tan\beta$, for fixed values 
of the charged, the CP--odd, and the fermiophobic Higgs masses.}
\label{cs_all_tb}
\end{figure} 
In Fig.~\ref{cs_all_tb} we plot all five mechanisms as a function of 
$\tan\beta$ for fixed values of the CP--odd Higgs mass $m_A=150$ GeV, 
charged Higgs mass $m_{H^+}=90$ GeV, and fermiophobic Higgs mass
$m_{h_f}=50$ GeV. For a fermiophobic Higgs of this mass to 
escape detection at LEP2 one requires $\tan\beta > 10$.
The traditional mechanism $p\overline p\to W^{\pm}h_f$
dominates at low $\tan\beta$ as expected, but falls fast with increasing
$\tan\beta$ due to the $\cos^2\beta$ suppression mentioned earlier. 
For $\tan\beta > 10$ all the new mechanisms offer larger 
cross--sections than the traditional one. The process 
$p\overline p\to H^{\pm}h_f$  
is dominant for $\tan\beta\gsim 3$ with a
cross--section growing from 30 fb for $\tan\beta=0.5$ up to 
155 fb for $\tan\beta=50$.
In the parameter space of interest ($\tan\beta\ge 10$)
one finds BR($H^{\pm}\to h_fW^{*})\approx 100\%$ and so
this mechanism essentially leads to a signature of $\gamma
\gamma\gamma\gamma$ plus $W^*$.

The $p\overline p\to H^+H^-$ mechanism has a production cross--section of
29 fb and is independent of $\tan\beta$. This cross--section becomes larger 
than $\sigma(p\overline p\to W^{\pm}h_f)$ at $\tan\beta\approx 7$,
and leads to a signature of $\gamma\gamma\gamma\gamma$ plus 
$W^*W^*$.
Similarly, $p\overline p\to H^{\pm}A^0$ has a cross--section
$\sigma=14$ fb (independent of $\tan\beta$) and supercedes the traditional
mechanism at $\tan\beta\approx 10$. As above, $h_f$ is produced via 
a cascade decay, which also provides the vector boson. 
%The final states $H^-A$ and $H^+A$ need to be summed over.
Both $H^{\pm}\to h_fW^{*}$ and $A^0\to h_fZ^{*}$ are 
effectively $100\%$ which leads again to the 
$\gamma\gamma\gamma\gamma$ plus $V^*V^*$ signature.
The behaviour of $\sigma(p\overline p\to A^0h_f)$ with $\tan\beta$ is similar
to that of $p\overline p\to H^{\pm}h_f$. It grows with $\tan\beta$ and is
essentially constant for large values of that parameter. This mechanism 
produces a fermiophobic Higgs directly, but has a lower rate due to 
the constraint $m_{h_f}+m_A\ge 200$ GeV. Since BR($A^0\to h_fZ)\approx 100\%$ 
the $\gamma\gamma\gamma\gamma$ plus $V^*$ signature also arises 
from this mechanism.

\begin{figure}
\centerline{\protect\hbox{\epsfig{file=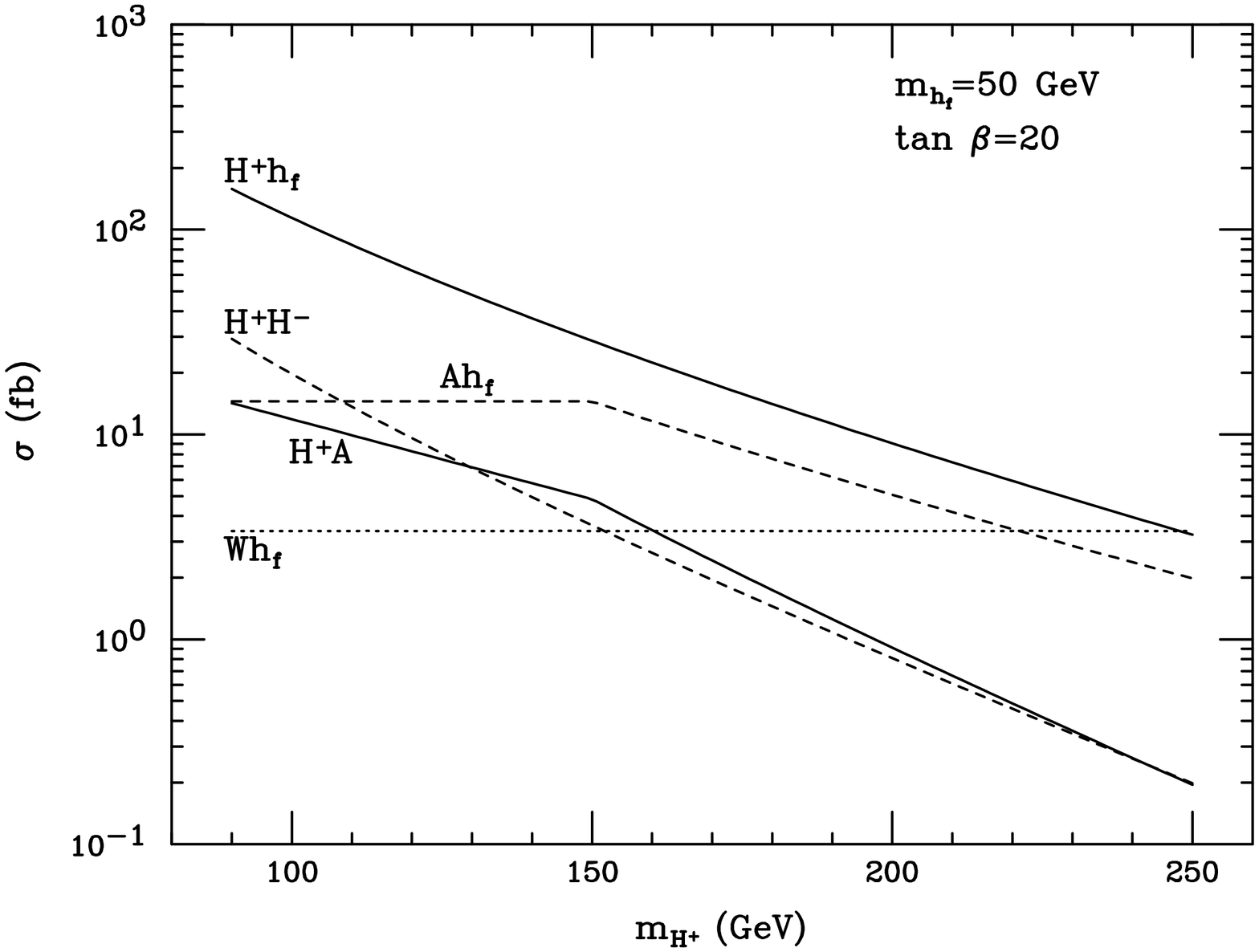,width=0.75\textwidth,angle=90}}}
\caption{\it Production cross section of five different modes leading to 
a fermiophobic Higgs boson as a function of the charged Higgs mass, for a 
fixed value of $\tan\beta$ and the fermiophobic Higgs mass (for the value 
of $m_A$, see the text).}
\label{cs_all_mH}
\end{figure} 
In Fig.~\ref{cs_all_mH} we plot the five mechanisms as a function of the 
charged Higgs mass $m_{H^+}$, for a constant value of the fermiophobic 
Higgs mass $m_{h_f}=50$ GeV. We also fix $\tan\beta=20$ which ensures that
a $h_f$ of this mass would have had too low a rate to be dectected
at LEP2 in the process $e^+e^-\to h_fZ$. In order to compare 
cross--sections that depend on $m_{H^+}$, like 
$\sigma(p\overline p\to H^+h_f)$, with cross--sections that depend 
on $m_A$, like $\sigma(p\overline p\to A^0h_f)$, we have taken $m_A=m_{H^+}$
provided $m_A>150$ GeV. Nevertheless, when $m_{H^+}<150$ GeV we keep a 
constant value $m_A=150$ GeV, which is required to satisfy 
the LEP constraint $m_{h_f}+m_A\ge 200$ GeV. This is
the explanation for the discontinuity in the slope of two of the 
cross--sections in Fig.~\ref{cs_all_mH}.
From the figure one can see that the traditional
mechanism $p\overline p\to W^{\pm}h_f$ is severely suppressed
($\sigma\approx 3$ fb and independent of $m_{H^+}$ and $m_A$), 
and there will not be enough events for its observation with the
expected Run IIa integrated luminosity of 2 fb$^{-1}$.

\begin{figure}
\centerline{\protect\hbox{\epsfig{file=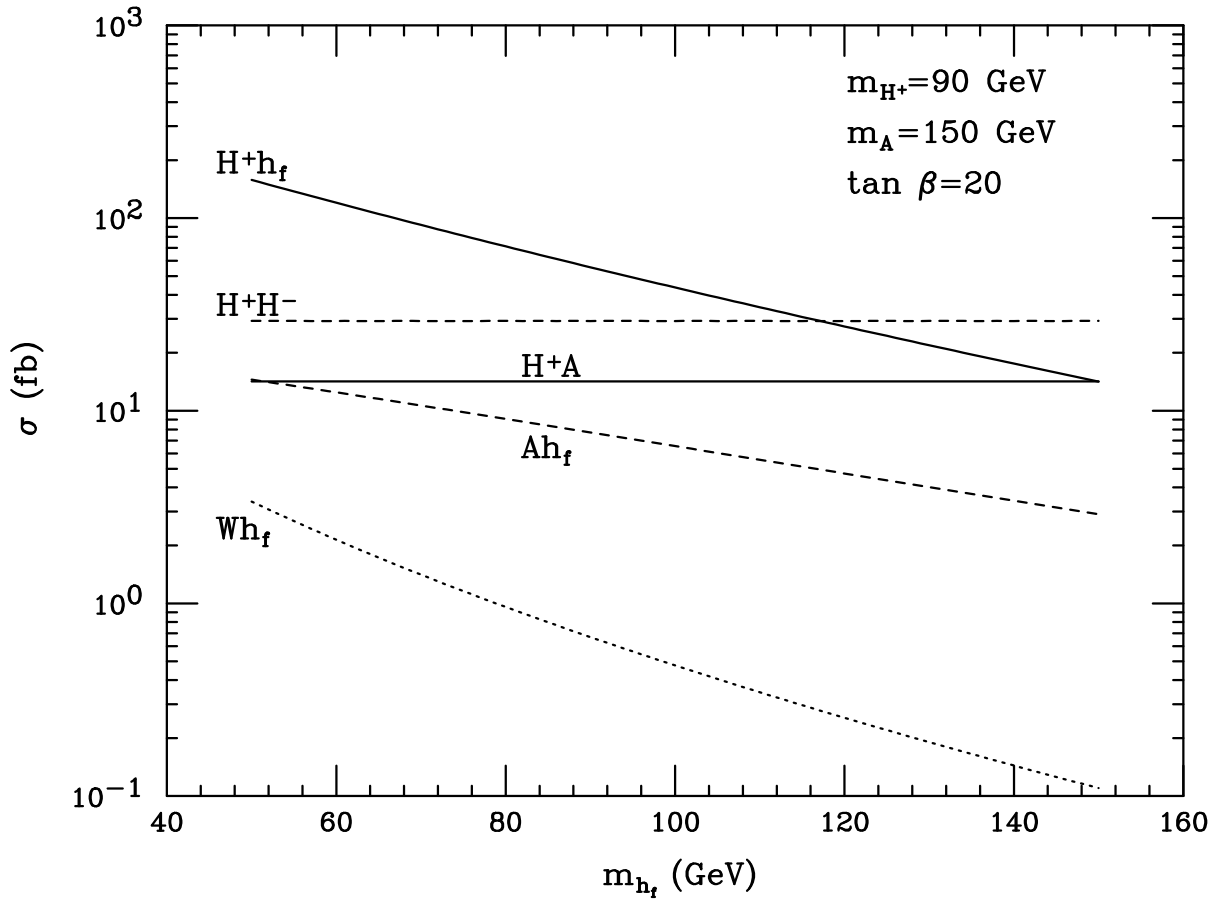,width=0.75\textwidth,angle=90}}}
\caption{\it Production cross--section of five different modes leading to 
a fermiophobic Higgs boson as a function of the fermiophobic Higgs mass, for 
fixed values of $\tan\beta$, the charged and the CP--odd Higgs masses.}
\label{cs_all_mf}
\end{figure} 

As in the previous case, the process with the highest cross--section is 
$p\overline p\to H^{\pm}h_f$. Due to phase space, this cross--section 
decreases with the charged Higgs mass from $\sigma\approx 160$ fb for
$m_{H^+}=90$ GeV to $\sigma\approx 3$ fb for $m_{H^+}=250$ GeV. Only
at these relatively high values of $m_{H^\pm}$ does
this cross--section become comparable with 
$\sigma(p\overline p\to W^{\pm}h_F)$. A
similar behaviour, although with a much smaller cross--section, is
also found for the process $p\overline p\to H^+H^-$, 
with a cross--section decreasing 
from $\sigma\approx 30$ fb for $m_{H^+}=90$ GeV to $\sigma\approx 0.2$ fb 
for $m_{H^+}=250$ GeV.
For most of the values of $m_{H^+}$ and $m_A$ shown in this graph, the 
subdominant mechanism is $p\overline p\to Ah_f$. Its cross--section 
remains constant at $\sigma\approx 15$ fb as long as $m_A=150$ GeV.
When $m_A$ increases, the cross section decreases to $\sigma\approx 2$ fb
for $m_A=250$ GeV, and thus it is observable only for 
lower values of $m_A$.
The last mechanism in this figure is $p\overline p\to H^+A$, and it
is only observable for $m_{H^+}\lsim 100$ GeV.

\begin{figure}
\centerline{\protect\hbox{\epsfig{file=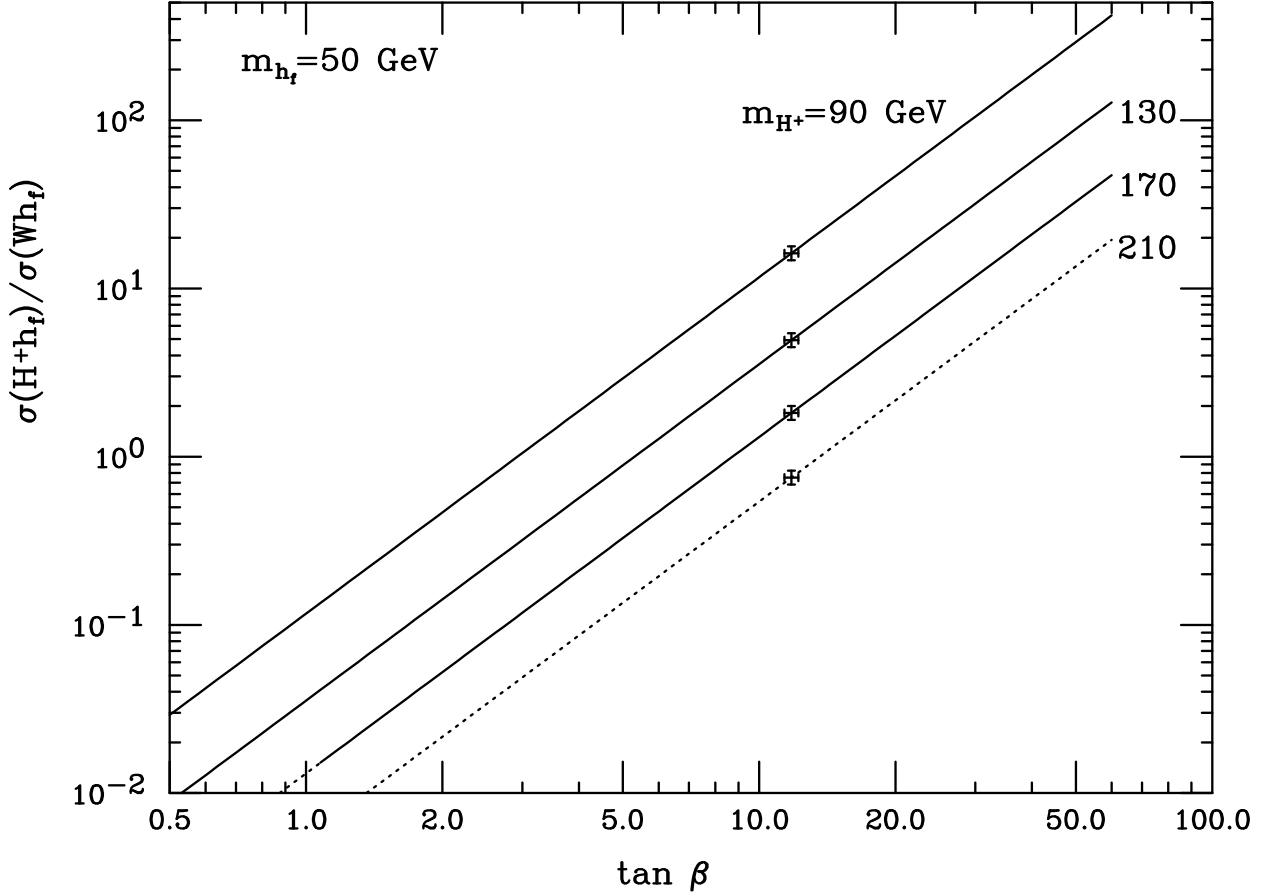,width=0.75\textwidth,angle=90}}}
\caption{\it Ratio of $\sigma(p\overline p\rightarrow H^{\pm}h_f$) and 
$\sigma(p\overline p\rightarrow W^{\pm}h_f$) as a function
of $\tan\beta$, for a fixed value of the fermiophobic Higgs mass and four
different values of the charged Higgs mass.}
\label{Hh_Wh_5}
\end{figure} 
In Fig.~\ref{cs_all_mf} we plot the cross--sections as a function of 
$m_{h_f}$.
For $m_{h_f}> 110$ GeV the decay $h_f\to VV^*$ becomes dominant. 
The figure shows the phase space suppression of increasing $m_{h_f}$.
The large value of $\tan\beta=20$ makes the traditional mechanism 
well suppressed and unobservable. The mechanism 
$p\overline p\to H^{\pm}h_f$ has a decreasing cross--section due to phase
space, and it is the most favourable for $m_{h_f}\lsim 120$. For larger
values of the fermiophobic Higgs mass, $p\overline p\to H^+H^-$ becomes
the largest cross--section (which is independent of $m_{h_f}$).
Note that for $m_{h_f}> m_{H^\pm}$ the mechanism 
$p\overline p\to H^{\pm}h_f$ does not provide a $V$ via the 
cascade decay of $H^\pm$, and thus only leads to a $\gamma\gamma$ signature.
In addition, $p\overline p\to H^{+}H^-$ would not produce a $h_f$
for $m_{h_f}> m_{H^+}$.

It is clear from the preceding figures that the production mechanism 
$p\overline p\to H^{\pm}h_f$ is usually 
the most favourable of the four alternatives we are analysing. 
Since all the mechanisms lead to the $\gamma\gamma V$ signature
one could in principle add all the cross--sections together. 
In the remaining two figures we compare more closely
$p\overline p\to H^{\pm}h_f$ and $p\overline p\to W^{\pm}h_f$.

In Fig.~\ref{Hh_Wh_5} we plot the ratio of 
$\sigma(p\overline p\to H^{\pm}h_f$)  
and $\sigma(p\overline p\rightarrow W^\pm h_f$)
as a function of $\tan\beta$. We take the fermiophobic Higgs 
mass $m_{h_f}=50$ GeV, and display four curves corresponding to different
charged Higgs masses. As in Fig.~\ref{cs_all_tb}, it is clear that the 
conventional production mechanism 
is convenient for low values of $\tan\beta$ and $p\overline p\to H^{\pm}h_f$ 
dominates for larger values of this parameter. 
The boundary lies somewhere between
$\tan\beta=3\to 13$ with the larger values obtained for large charged Higgs 
masses. The cross on each curve marks the threshold of observability
(which we take as 10 fb) for $\sigma(p\overline p\to W^{\pm}h_f)$, 
and corresponds to $\tan\beta\approx 12$. To the left of the crosses 
$\sigma(p\overline p\to W^{\pm}h_f)> 10$ fb, and to the right 
$\sigma(p\overline p\to W^{\pm}h_f)< 10$ fb. 
The solid lines correspond to 
$\sigma(p\overline p\to H^{\pm}h_f)\ge10$ fb
and dotted lines to $\sigma(p\overline p\to H^{\pm}h_f)\le10$ fb.
As observed in Fig.~\ref{cs_all_tb}, 
$\sigma(p\overline p\to H^{\pm}h_f)$ grows fast with $\tan\beta$, until
it saturates at around $\tan\beta\approx5$. This saturation value is
$\sigma(p\overline p\to H^{\pm}h_f)=158$, 48, 18, and 7 fb for 
$m_{H^{\pm}}=90$, 130, 170, and 210 GeV respectively (the last one being
unobservable).

\begin{figure}
\centerline{\protect\hbox{\epsfig{file=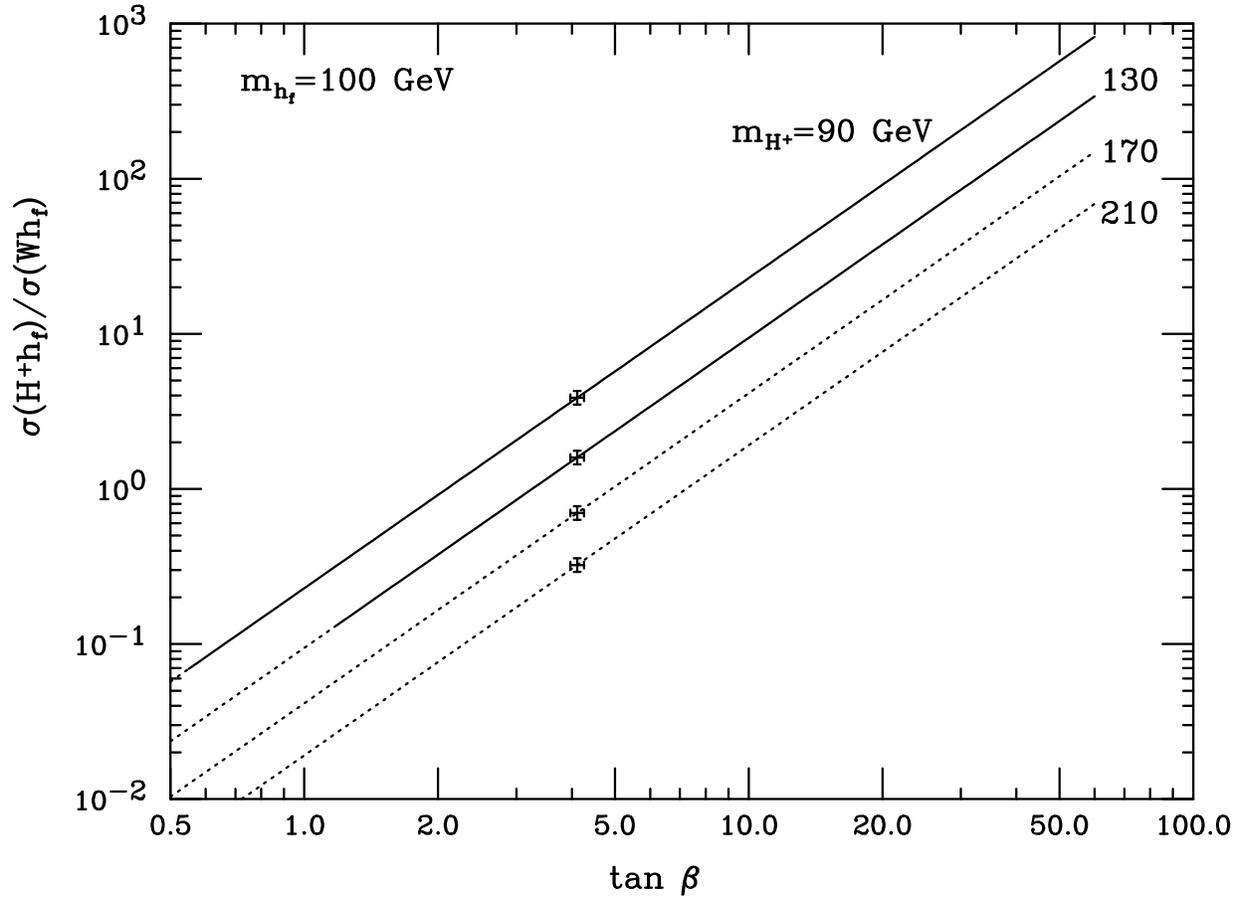,width=0.75\textwidth,angle=90}}}
\caption{\it Ratio of $\sigma(p\overline p\rightarrow H^{\pm}h_f$) and 
$\sigma(p\overline p\rightarrow W^{\pm}h_f)$ as a function
of $\tan\beta$, for a fixed value of the fermiophobic Higgs mass and four
different values of the charged Higgs mass.}
\label{Hh_Wh_10}
\end{figure} 
Fig.~\ref{Hh_Wh_10} is the same as Fig.~\ref{Hh_Wh_5} 
but with $m_{h_f}=100$ GeV. Here the conventional
mechanism is unobservable for $\tan\beta\gsim4$. 
The saturation values are
$\sigma(p\overline p\to H^{\pm}h_f)=44$, 18, 8, and 4 fb for the same values 
of the charged Higgs mass.
The new mechanism overcomes the conventional one in 
a larger region of parameter space since the ratio of cross--sections is
larger than one for $\tan\beta\gsim 2\to 7$, depending on the 
charged Higgs mass.

Given the sizeable cross--sections for $p\overline p\to H^{\pm}h_f$
this process (with $h_f$ replaced by $h^0$) might have a wider application, 
e.g. in the search for $H^\pm$ of any (non--SUSY) 2HDM. In particular 
this process is maximized in the parameter space of a 
light $h^0$ with suppressed couplings to vector bosons 
(i.e. small $\sin(\beta-\alpha)$).
%This mechanism probes larger values of
%$m_{H^\pm}$ than $p\overline p\to H^+H^-$ and 
%$p\overline p\to t\overline t$, $t\to H^\pm b$, 

\section{Conclusions}
We proposed new production mechanisms for light fermiophobic Higgs
bosons ($h_f$) with suppressed couplings to vector bosons ($V$)
at the Fermilab Tevatron.
Importantly the new mechanisms offer sizeable cross--sections
when the conventional process $(qq'\to W^\pm h_f$) 
is suppressed, and provide distinctive signatures with up to
4 photons. We showed that $qq'\to H^\pm h_f$ is particularly promising
with cross--sections as large as 150 fb if both $h_f$ and $H^\pm$ are
light $(< 100$ GeV). 
We suggested that the mechanism $qq'\to H^\pm h_f$ might also have
a wider application in the search for a 
light $h^0$ and $H^\pm$ of any general (non--SUSY) 2HDM.

\section*{Acknowledgements}  
A.G.A. expresses his gratitude to 
La Universidad Cat\'olica de Chile where part of this work was carried out.
M.A.D. is thankful to Korea Institute for Advanced Study for their kind 
hospitality. Part of this work was financed by CONICYT grant No.~1010974.

\end{document}